\documentclass[pre,floatfix,twocolumn,noshowpacs,superscriptaddress]{revtex4-1}

\usepackage[english]{babel}
\usepackage[dvips]{graphicx}
\usepackage{multirow,booktabs,longtable}
\usepackage{color}
\usepackage{amssymb,amsfonts,amsmath}

\begin{document}

\title{World citation and collaboration networks: uncovering the role of geography in science}

\author{Raj Kumar Pan}
\affiliation{Department of Biomedical Engineering and Computational Science, Aalto University School of Science, P.O.  Box 12200, FI-00076, Finland} 
\author{Kimmo Kaski}
\affiliation{Department of Biomedical Engineering and Computational Science, Aalto University School of Science, P.O.  Box 12200, FI-00076, Finland} 
\author{Santo Fortunato}
\affiliation{Department of Biomedical Engineering and Computational Science, Aalto University School of Science, P.O.  Box 12200, FI-00076, Finland} 
\affiliation{Complex Networks and Systems Lagrange Laboratory, Institute for Scientific Interchange (ISI), Torino, Italy}

\begin{abstract}
  Modern information and communication technologies, especially the
  Internet, have diminished the role of spatial distances and territorial
  boundaries on the access and transmissibility of information. This has
  enabled scientists for closer collaboration and internationalization.
  Nevertheless, geography remains an important factor affecting the
  dynamics of science. Here we present a systematic analysis of citation
  and collaboration networks between cities and countries, by assigning
  papers to the geographic locations of their authors' affiliations. The
  citation flows as well as the collaboration strengths between cities
  decrease with the distance between them and follow gravity laws. In
  addition, the total research impact of a country grows linearly with the
  amount of national funding for research \& development. However, the
  average impact reveals a peculiar threshold effect: the scientific output
  of a country may reach an impact larger than the world average only if
  the country invests more than about 100,000 USD per researcher annually.
\end{abstract}

\maketitle

\section{Introduction}
The strength of most interactions in nature typically decreases with
the distance between objects or constituents. The most famous example is Newton's
gravitational force, which is known to decay with the square of the
distance between the masses. This principle holds also outside the realm of
physical processes. Recent studies on mobile phone communication
networks~\cite{Lambiotte08,Krings09} and blogs~\cite{LibenNowell05} have
revealed that the probability for a social tie to occur between agents
decays with a power of their distance.

Likewise, scientific interactions are likely to take place between
scholars localized in the same or nearby areas. Scientists tend to cluster
in space, since the elaboration and progress of a project requires
frequent discussions between collaborators that is hardly possible
if they live far apart. Factors based on cultural, linguistic and
institutional differences cause additional obstacles to
long-distance cooperation~\cite{Okubo04}. Further, research funding is mostly
allocated at the national level~\cite{banchoff02}, thus favoring regional
over international collaborations.

Nowadays, the Internet and the greater affordability of international
transportation have enormously reduced distances between people, overcoming
both geographic and cultural barriers~\cite{Cairncross01,Finholt97,
Teasley01}.  This in turn has made scientific collaborations between
distant scholars far easier than
before~\cite{Georghiou98,Rosenblat04,Havemann06, Chandra07, Agrawal08,
Hennemann12}. Nevertheless, the role of geography in the creation
and recognition of scientific output is not yet fully known. For example,
How do scientific interactions depend on distance?
Is collaboration concentrated within the perimeter of a university, of a
city or of a country, as it used to be in the past,
or has it become truly international, possibly due to the modern
information and communication technologies?

Multi-authored collaborations serve as big opportunity for
science~\cite{katz97}, as one can integrate a wide range of
competence and skill, to attack difficult problems,
with an enhanced chance of success. Indeed, the last
decades have witnessed the formation of larger and larger research
teams~\cite{Adams05,Wuchty07}. In particular, multi-university
collaborations have been growing at a fast pace and are more likely to lead
to high impact publications~\cite{Jones08}, especially if they involve
different countries~\cite{narin91,glanzel99}. On the other hand, there
is also evidence of decreasing returns from large team size, likely from
management inefficiencies, which limits the productivity arising from
collaboration~\cite{Petersen12}.

Geographic proximity is also likely to affect the process of giving and
receiving credits for someone's work, expressed by paper
citations. For most papers one expects
to find a decaying probability of citation with distance, as new findings
are typically more visible in the area where the authors operate. This is
confirmed by a recent study~\cite{Borner06}. In addition, collaboration
patterns are likely to influence and be influenced by citations. While
collaborating, scholars become more familiar with the scientific output of
their co-authors, which then has a higher chance to be cited in the future.
In turn, scholars citing frequently each other's work have strongly
overlapping research interests, and are more likely to become co-authors sooner
or later. Therefore citations and collaborations between distinct locations
are likely to be correlated. 
 However, it is crucial to assess how
  collaborative patterns affect citation flows, to be able to disentangle
  the actual impact of a publication (and, therefore, its merit)
  from credits coming through social networking. A geographic analysis of
  citation flows between cities is also useful to understand how quickly a
  new result gets recognized by the scientific community in different
  geographical areas, which may help to uncover how new scientific
  paradigms spread and get established~\cite{Pan12SR}.

Knowing how scientific interactions vary with distance is also valuable for
practical reasons. To scholars, it might suggest how to choose
collaborators in order to optimize the impact and visibility of their
research.  To institutions and governments, it might advice suitable
allocations of funds for regional and international projects, in order to
improve the scientific outcome for a given amount of resources.  It is then
not surprising that spatial scientometrics has acquired a prominent role
during the last few years. There are a number of studies carried out
exploiting the enhanced availability of citation data~\cite{frenken09}. 
Yet there are other factors, namely funding, that also plays a crucial role
in the development of a research project, as it not only contribute towards
the direct and overhead costs of the research but also facilitates the
cooperation and collaboration among researchers working in different
locations and different fields~\cite{Lee05}. Since both public and
industrial resources are used to fund academic research, it is also natural
to question the result and impact obtained with these
resources~\cite{Arora98,Arora05}.

We have performed the first comprehensive study of citation and
collaborative interactions between different geographic locations. We used
one of the world's largest citation databases to derive the citation and
the collaboration network, i.e. weighted networks where nodes are cities
and links are citations and collaborations between the corresponding cities
(see Methods). The analysis of these networks~\cite{caldarelli07,barrat08,Newman10,rosvall10} discloses the
existence of gravity laws as well as non-trivial correlation between
collaborations and citations.  Finally, we explore the issue
of the importance of funding to research and development in promoting high
quality science, by studying the relationship between national expenditure,
the number of publications and their impact in terms of number of
citations for different countries.
%We also present a global study of the relationship between funding and
%scientific outcome of different countries.

\section{Results}
\begin{figure*}[tb]
  \centering
  {\includegraphics[width=0.80\linewidth]{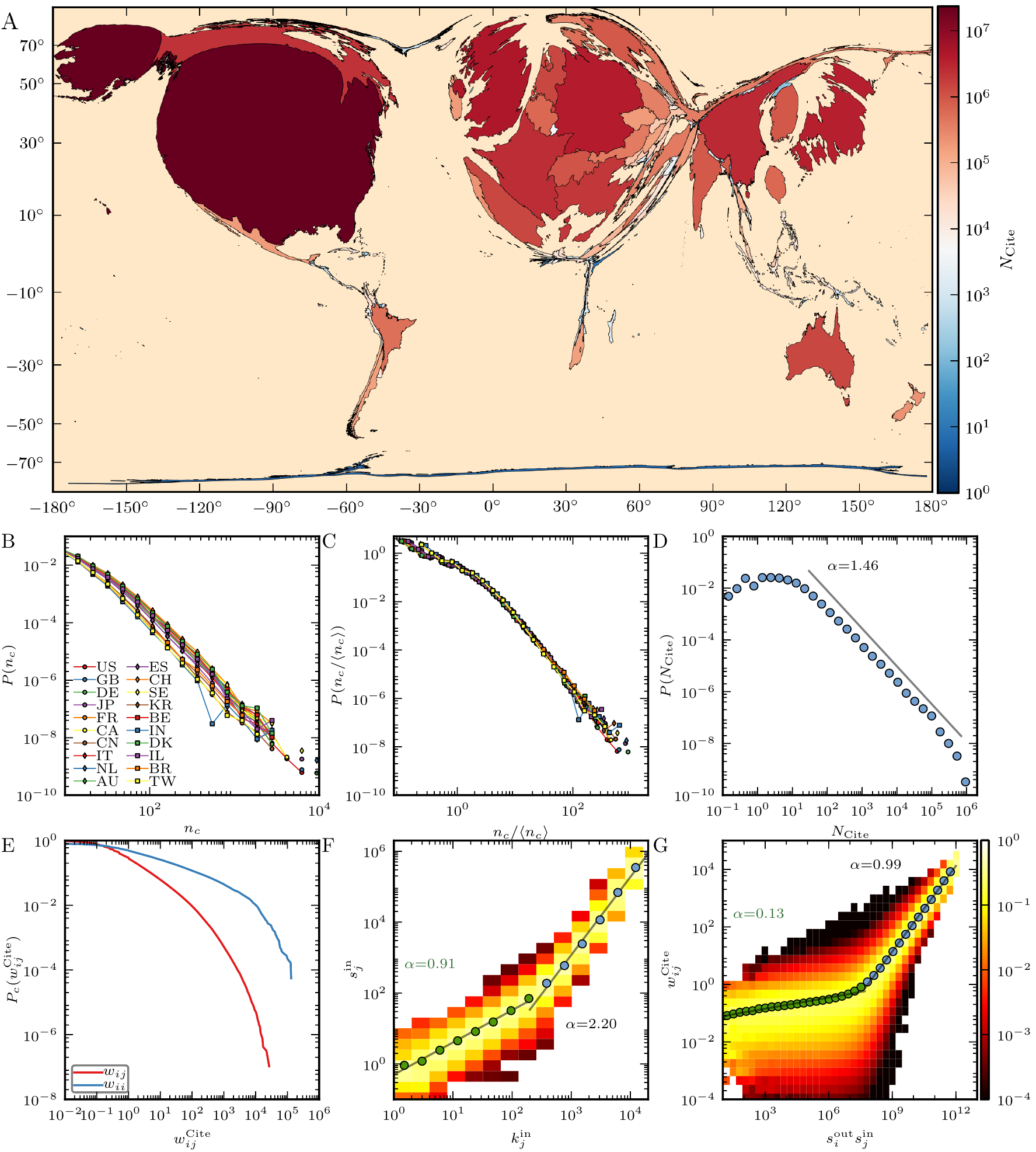}}
\caption{Properties of the world citation network. (A) Citation map of the
  world where the area of each country is scaled and deformed according to
  the number of citations received, which is also represented by the
  color of each country. 
(B) Citation distribution of
  papers of top 20 countries. If a paper is written by authors from
  multiple countries,
  the paper contributes to each country. (C) When the distributions in
  (B) are normalized by the average number of citations of each
  country, they fall on top of each other. (D) Probability distribution
  function of the number of citations received by each city. (E) Cumulative
  distribution function of the link weights $w_{ij}$ (excluding self-links)
  and self-links $w_{ii}$ in the citation network of cities. (F) Node
  in-strength against its in-degree for the city citation network. (G) Link weight against the product of the strengths of the
  connected nodes in the city citation network. For each plot we show
the corresponding best-fit lines and power law exponents.}
\label{fig:citation}
\end{figure*}

The research contribution of each country in terms of the (normalized)
number of citations received $N_{\mathrm{Cite}}$ is illustrated in the
world map of Fig.~\ref{fig:citation}A. Colored maps can be misleading as
the value assigned to a large area gives an impression of a much greater
impact of that color in the visualization. We thus
created a cartogram, in which the geographic regions are deformed and
rescaled in proportion to their relative research
contribution~\cite{Gastner04}. The citation strengths of countries span
over seven orders of magnitude. North America and
Europe receive 42.3\% and 35.3\% of world's citations, respectively.
In contrast, the contribution by Asia amounts to only 17.7\% of world's
citations while the total contribution of Africa, South America and Oceania is
lower than 5\%. In this ranking the United States is the leading
country followed by the United Kingdom, Germany, Japan, and China. 
The corresponding world map in terms of countries' number of
(normalized) publications is shown in the Appendix Fig.~\ref{fig:citationSI}.
 This heterogeneity suggests that a small number of countries have a
  substantial contribution to research while the rest has
a negligible contribution. In Appendix Fig.~\ref{fig:averageCitationSI} we report the results
for the average number of citations of each country. 

In order to find
out the quality of papers published by different countries we consider
the number of citations of each of the papers written by that country.
In Fig.~\ref{fig:citation}B we plot the probability distribution of the
number of citations of papers in the largest 20 countries. A paper is
associated to a country if at least one of its affiliations is from that
country. All these distributions are broad and vary over four orders of
magnitude. When each distribution is rescaled by the average number of
citations of papers of the respective country, all curves nicely
collapse (Fig.~\ref{fig:citation}C).
This result suggests that the functional form of the citation distribution
is the same in each country and that the difference between countries can
be effectively summarized by the average number of citations. This type of
universality holds at the level of scientific disciplines as
well~\cite{radicchi08}. 
%Supplementary Fig.~S2 shows the world map in terms of the average
%number of citations per paper of each country.

Next we consider the contribution at the level of cities. In
Fig.~\ref{fig:citation}D we plot the probability distribution of the
cities' citations. The distribution is broad, spanning over five orders of
magnitude, and it follows a power law decay with exponent
$1.46\pm0.03$. 
This suggests a relationship with the population of the city,
as the city size distribution obeys the Zipf
law~\cite{Zipf49, Gabaix99}, i.e. decays as a power law (with
exponent $2$).
 The observed power law scaling relation might suggest a
  self-organization phenomena due to the agglomeration benefits in science.
  These advantages can be due to the ease in collaboration between groups
  working in similar fields, sharing of infrastructure and support, etc.,
  which leads to efficient integration and transfer of information.

We now consider the weighted citation network between cities, where the
nodes are the cities that are connected by weighted and directed links,
indicating publications of one city citing publications of the others. The network has 18,199 nodes and 9,494,021 links including 14,447
self-links (i.e., citations within the same city). In
Fig.~\ref{fig:citation}D we plot the cumulative distribution of
the weights of self-links and links between different nodes. Both these
distributions are broad; however, the weights of self-links are more
heterogeneous, revealing a bias towards self-citations.
Next we calculate the number of incoming links, i.e., the in-degree
$k_i^{\mathrm{in}}$ of each node $i$ and its in-strength,
$s_i^{\mathrm{in}}=\sum_j w_{ji}^{\mathrm{Cite}}$, which equals the
number $N_i^{\mathrm{Cite}}$ of (normalized) citations received.  By plotting the
in-degree against the in-strength, we find that there is a power law scaling
behavior with $\langle s^{\mathrm{in}} \rangle (k^{\mathrm{in}}) \propto
(k^{\mathrm{in}})^{\alpha}$
(Fig.~\ref{fig:citation}E). 
However, there are two distinct scaling
regimes: for nodes with small $k_i^{\mathrm{in}}$ ($<200$) the exponent is
$\alpha=0.91\pm0.03$ ({\text regression coefficient $\pm$ standard error of the
estimate $R=0.95\pm0.01$}), while for large
$k_i^{\mathrm{in}}$ ($\geq200$) the exponent is $\alpha=2.20\pm0.08$
($R=2.01\pm0.01$). The super-linear behavior suggests that stronger links are more
frequently connected to high in-degree nodes. The out-strength of the nodes
follows a similar relationship with the out-degree of the nodes (see
Appendix Fig.~\ref{fig:citationSI}). 
Finally, we plot the weights of the links $w_{ij}^{\mathrm{Cite}}$ against
the product of the node strength $s_i^{\mathrm{out}}s_j^{\mathrm{in}}$. The
product $s_i^{\mathrm{out}}s_j^{\mathrm{in}}$ gives the weight of a link
that is expected to occur by chance between $i$ and $j$ if all the papers would be
citing each other at random. Even in this case there are two distinct
scaling regions, $w_{ij}^{\mathrm{Cite}} \propto
(s_i^{\mathrm{out}}s_j^{\mathrm{in}})^{\alpha}$, where $\alpha=0.13\pm0.01$
($R=0.19\pm0.0003$) if the product is less than $2\times10^7$, while for larger
values of the product $\alpha=0.99\pm0.01$ ($R=1.07\pm0.001$). This suggests that the observed citation is as expected
between high strength nodes, while it is much lower in case of cities with
low strength.

\begin{figure}
  \centering
 {\includegraphics[]{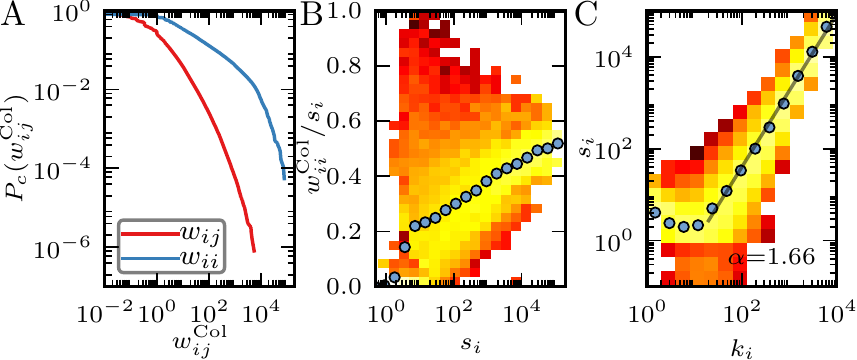}}
\caption{Properties of the world collaboration network. (A) Cumulative probability
  distribution of the link weights in the collaboration network of cities.
  Self-links are shown separately. (B) Fraction of internal collaboration,
  indicated by the ratio of the weight $w_{ii}^{\mathrm{Col}}$ of the
  self-link and strength $s_i$ of a node, against $s_i$. (C) Strength of a
  node against its degree. The straight line indicates a power law behavior
  with exponent $1.66\pm0.04$. In these plots we use the same colorbar as
in Fig.~1.}
\label{fig:collaboration}
\end{figure}
Let us now consider the collaboration network at the city level, where the
nodes are cities and weighted undirected links indicate the presence and
frequency of collaborations between scholars of different cities. There are
18,199 nodes in the network and 1,256,718 undirected links including
14,954 self-links. The weight of the self-links indicates the amount of
internal collaboration. 
The degree of a node $i$ indicates the number
of other cities with which $i$ collaborates and its strength is
indicative of, but not coincident with,
the number of papers written by scholars of institutions in that
city. 

In Fig.~\ref{fig:collaboration}A we plot
the cumulative probability distribution of link weights. As for
citations, the weights of self-links are more broadly distributed than the
weights of the links between different cities, showing that scholars of a
city collaborate more frequently with each other than with colleagues from
any other city. The distributions of collaboration and citation
streams between cities differ from their analogues in
mobile phone communications and world trade, that show log-normal
distributions~\cite{Krings09,Bhattacharya08}.
Next, we consider the fraction of internal collaboration by calculating the
ratio of the weight of the self-link to the strength of the node. By plotting
$w_{ii}^{\mathrm{Col}}/s_i$ against the strength of the node $s_i$, we see
that the ratio increases with $s_i$, indicating that as the city size
increases most of its collaborations take place within the city (Fig.~\ref{fig:collaboration}B).  However,
for small cities most of their papers are written with external collaborators. The
node degree scales with its strength as $\langle s \rangle (k) \propto
k^{\alpha}$, where $\alpha=1.66\pm0.04$ ($R=1.65\pm0.01$) (Fig.~\ref{fig:collaboration}C). This super-linear scaling suggests that higher
degree nodes are more frequently connected by stronger links. 

\begin{figure}
  \centering
  {\includegraphics[]{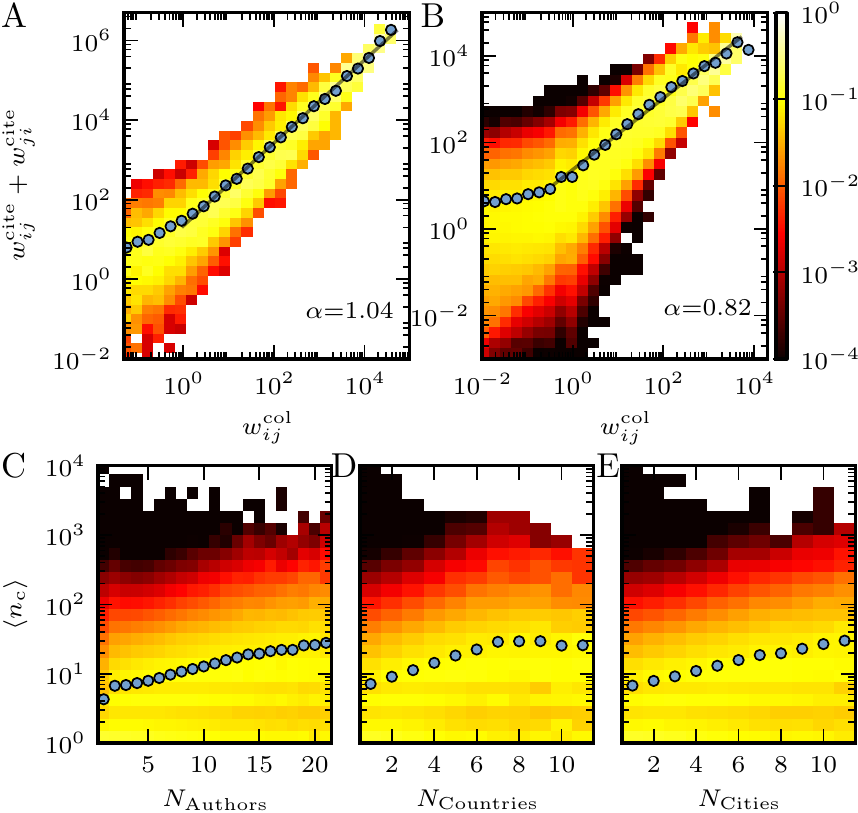}}
\caption{Correlation between the world citation and collaboration networks.
  Weight of the links in the citation network against the corresponding
  links in the collaboration network at the (A) country level and (B) city
  level network. Power law scaling is shown by solid lines with
  exponents $1.04\pm0.01$ and $0.82\pm0.02$, respectively. Density plot of
  the number of citations of a publication against the number of (C)
  co-authors, (D) countries (E) cities in the affiliation. The circles
  indicate the average trend.}
\label{fig:collaborationAndCitation}
\end{figure}
Let us explore the relationship between the citation and the
collaboration networks at both the country and the city level. At the country
level the collaboration network comprises 226 nodes and 10,308
undirected links, including 219 self-links. In the citation network there
are also 226 nodes but 28,869 directed links, including 215 self-links.
In Fig.~\ref{fig:collaborationAndCitation}, we plot the weight of links of
the collaboration network, $w_{ij}^{\mathrm{Col}}$ against the weight of
the same links in the citation network, $w_{ij}^{\mathrm{Cite}}+w_{ji}^{\mathrm{Cite}}$. We find
scaling $w_{ij}^{\mathrm{Col}} \propto (w_{ij}^{\mathrm{Cite}}+w_{ji}^{\mathrm{Cite}})^{\alpha}$
where $\alpha=1.04\pm0.01$ ($R=1.08\pm0.008$) for countries (Fig.~\ref{fig:collaborationAndCitation}A), and
$\alpha=0.82\pm0.02$ ($R=1.05\pm0.002$) for cities
(Fig.~\ref{fig:collaborationAndCitation}B), 
i.e. the increase in collaboration is linearly related to the amount
of citations exchanged between the two countries/cities. 

\begin{table}
\centering
\caption{Dependence of citations on collaboration. We categorize each paper
by the number of authors and their affiliations. For each of these groups
we indicate the fraction of papers that are in the group and the
mean number of citations. The error represents the standard error of
the mean,
calculated using bootstrap sampling with repetition.}
\label{tab:nAuthors}
%\small
%\tabcolsep=0.11cm
\begin{tabular}{ccccc}
  \hline
$N_{\mathrm{Authors}}$ & $f_{\mathrm{Papers}}$ &
Single & Multiple & Multiple \\ 
          & (in \%) & City & City & Countries \\ 
\hline
1     & 13.03 & 4.25 $\pm$ 0.02 & 4.95 $\pm$ 0.12 & 5.24 $\pm$ 0.11    \\ 
2     & 19.01 & 6.80 $\pm$ 0.02 & 6.11 $\pm$ 0.04 & 7.00 $\pm$ 0.05    \\ 
3     & 18.34 & 6.92 $\pm$ 0.02 & 6.38 $\pm$ 0.03 & 7.30 $\pm$ 0.04    \\ 
4     & 14.95 & 7.19 $\pm$ 0.02 & 7.02 $\pm$ 0.03 & 8.03 $\pm$ 0.04    \\ 
5     & 11.10 & 7.62 $\pm$ 0.03 & 7.66 $\pm$ 0.03 & 8.79 $\pm$ 0.04    \\ 
6     & 8.01  & 8.13 $\pm$ 0.04 & 8.52 $\pm$ 0.05 & 9.77 $\pm$ 0.05    \\ 
7     & 5.20  & 8.85 $\pm$ 0.05 & 9.56 $\pm$ 0.07 & 10.90 $\pm$ 0.07   \\ 
8     & 3.45  & 9.50 $\pm$ 0.07 & 10.67 $\pm$ 0.09 & 12.10 $\pm$ 0.10  \\ 
9     & 2.22  & 10.23 $\pm$ 0.10 & 11.52 $\pm$ 0.12 & 13.17 $\pm$ 0.12 \\ 
10    & 1.53  & 10.57 $\pm$ 0.12 & 12.45 $\pm$ 0.14 & 14.70 $\pm$ 0.15 \\ 
$>$10 & 3.17  & 13.82 $\pm$ 0.17 & 16.64 $\pm$ 0.16 & 21.37 $\pm$ 0.17 \\ 
 \hline
\end{tabular}
\end{table}
We now consider the dependence of the number of
citations of a paper on the number of coauthors of that paper and on the
number of affiliations of its coauthors.
It has been previously shown that papers published by teams often
get more citations than single author papers~\cite{Wuchty07,Jones08}. Our
results also show that the average number of cites of a publication
increases with the number of co-authors of that
publication~(Fig.~\ref{fig:collaborationAndCitation}C). Furthermore, the average number of citations
of a publication increases with the number of affiliated countries and
cities of its authors (Fig.~\ref{fig:collaborationAndCitation}D and E).
In order to separate the effect of the number of coauthors and
  different type of collaboration (internal, domestic and
  international) we grouped each paper based on its affiliations and number
of coauthors.
In Table~\ref{tab:nAuthors}, we consider papers with a given
  number of authors and categorize them according to whether all the
  affiliations listed in the paper are from a single city, from multiple
  cities in a single country or from different countries. For an equal
  number of authors, publications having multiple
  international affiliations get a statistically significant increment ($p<10^{-4}$)
  in the number of citations with respect to
  publications with only domestic affiliations. 
Thus, crossing territorial boundaries also pays off in terms of scientific
impact. In contrast,  multiple domestic affiliation do not positively
effect the number of citations when the number of authors in a publication
is less than $6$.

\begin{figure}[t]
  \centering
  \includegraphics[width=1.0\linewidth]{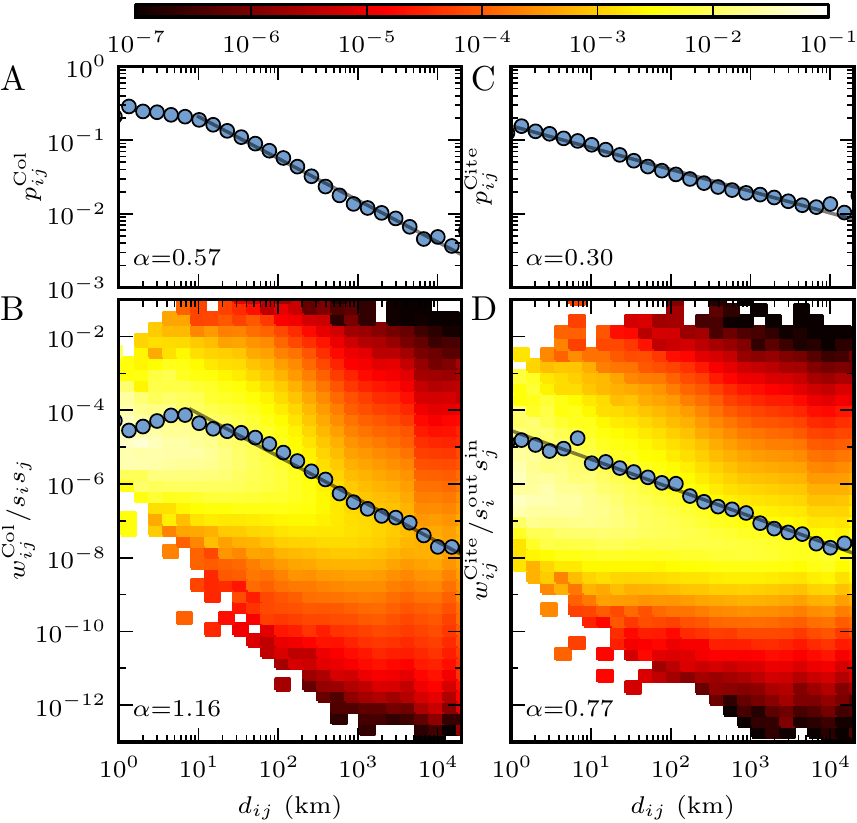}
\caption{Effect of geographical proximity in the world collaboration and
  citation networks. The  probability of existence of a link as a function
  of the distance between two cities in the (A) collaboration network and
  (B) citation network. Distribution of the ratio of
  the link weight and product of the strengths of its endpoints in
  (C) collaboration network, $w_{ij}^{\mathrm{Col}}/s_is_j$ and (D)
  citation network,
  $w_{ij}^{\mathrm{Cite}}/s_i^{\mathrm{out}}s_j^{\mathrm{in}}$ against the
  distance $d_{ij}$ between the cities. For each distance the average ratio
  is also shown. The solid line indicates a power law behavior with
  exponent $\alpha=1.16\pm0.03$ and $0.77\pm0.02$ respectively.}
\label{fig:gravityCity}
\end{figure}
Next we consider the effect of geographical proximity on the citation and
collaboration networks by determining the geographic location (latitude and
longitude) of each place in the dataset~\cite{Barthelemy11} (see Methods). We found that the probability that there is a link between
two cities in the collaboration network decreases as a power law as the
distance between the two cities increases (Fig.~\ref{fig:gravityCity}A).
The power law exponent is $0.57\pm0.01$. Our results are different from
those obtained in Ref~\cite{Katz94}, where it was found that the
distribution of distances between co-authors decreases exponentially. Such
difference might be due to the limited dataset used in
Ref~\cite{Katz94}, which included only papers published before 1990,
and possibly also due to the recent advances in communication and
transportation technologies. 

Many spatially embedded networks have been observed to follow
gravity laws~\cite{Barthelemy11}, where the flow between two locations
follows
\begin{equation}
  T_{ij} \propto \frac{P_iP_j}{d_{ij}^\alpha}. 
  \label{eq:gravity}
\end{equation}
Here, $T_{ij}$ is the flow between nodes $i$ and $j$, $P_i$ and $P_j$ are
the populations of nodes $i$ and $j$, respectively and $d_{ij}$ is the
geodesic distance between $i$ and $j$, the value of exponent $\alpha$
being dependent of the system. For the collaboration network
Eq.~\ref{eq:gravity} becomes
\begin{equation}
  w_{ij}^{\mathrm{Col}} \propto
  \frac{s_is_j}{d_{ij}^{\alpha}}.
  \label{eq:gravityCollaboration}
\end{equation}
In Fig.~\ref{fig:gravityCity}B, we plot the ratio
${w_{ij}^{\mathrm{Col}}}/({s_i s_j})$ against the distance $d_{ij}$ between all node
pairs. We found that as the distance increases 
$\langle {w_{ij}^{\mathrm{Col}}}/({s_i s_j}) \rangle$ decreases as a power
law with the exponent $\alpha=1.16\pm0.03$ ($R=-0.97\pm0.002$), except at very
short distances. 
As we have seen
before, collaboration and citation between two places are
correlated. Hence, we also look at the geographical proximity in the
citation network.  We found that the probability that there is a link
between two cities in the citation network also decreases with distance as
a power law (Fig.~\ref{fig:gravityCity}C). In this case the power law exponent is much lower ($0.30\pm0.01$). 
The gravity law for the citation network reads
\begin{equation}
  w_{ij}^{\mathrm{Cite}} \propto
  \frac{s_i^{\mathrm{out}}s_j^{\mathrm{in}}}{d_{ij}^{\alpha}}.
  \label{eq:gravityCitation}
\end{equation}
In Fig.~\ref{fig:gravityCity}D we plot
${w_{ij}^{\mathrm{Cite}}}/({s_i^{\mathrm{out}}s_j^{\mathrm{in}}})$ against the
distance between all the node pairs in the citation network. As for
the collaboration network we found that 
$\langle {w_{ij}^{\mathrm{Cite}}}/({s_i^{\mathrm{out}}s_j^{\mathrm{in}}})
\rangle$ decreases with distance as a
power law with the exponent $\alpha=0.77\pm0.02$ ($R=-0.35\pm0.001$). The above
analysis shows the existence of an important spatial component in both the
citation and the collaboration network. It shows that both our
collaborators and our citations typically come from our spatial neighborhood. Further,
long distance collaborations as well as citations decrease as a power law
of distance. The difference of the scaling exponents of the two networks
suggests that two distant places are more likely to cite each other than
collaborate.
Additional results are shown in the Appendix Fig.~\ref{fig:gravityLawSI}.

\begin{figure}
  \centering
  {\includegraphics[width=1.0\linewidth]{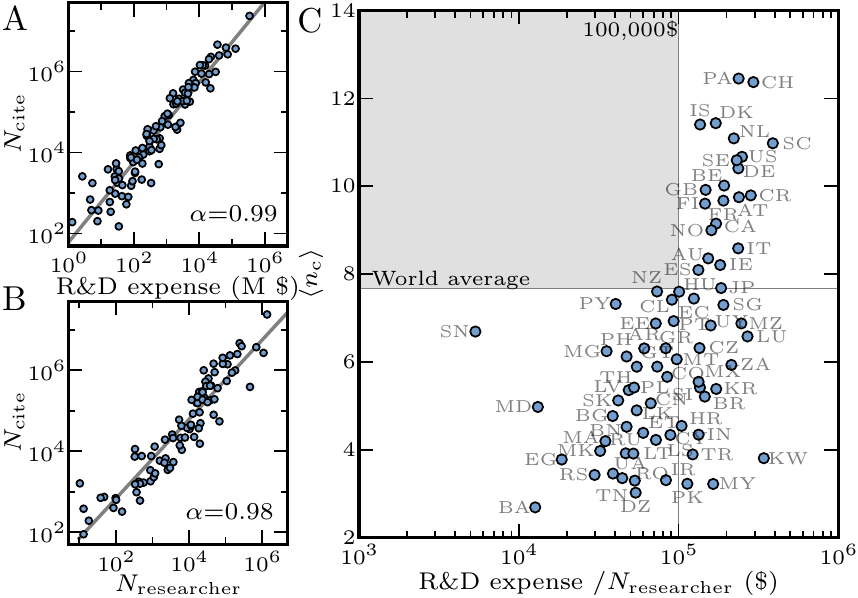}}
\caption{Relation between research outcome and funding.  Average
  number of citations per paper of a country against (A) the expenditure in research and
  development (in millions of dollars per year, and purchasing power parity)
  and (B) the number of researchers in that country. The solid line
  indicates power law scaling with exponent $0.99\pm0.03$ and $0.98\pm0.04$,
  respectively. (C) Average number of citations per paper of a country against the
  average spending per researcher. The horizontal line indicates the
average number of citations over all papers of all countries, the vertical
line indicates the threshold of about 100,000 \$ per researcher per year.}
\label{fig:gdp}
\end{figure}
The research performance of each country is generally estimated on the
basis of the number of publications and citations. Although these are
straightforward measurements of research output, they depend on a wide
spectrum of resources~\cite{Johnes95}. For instance, the number of
researchers and facilities (instruments, laboratories, libraries and other
resources) available are typically different in different countries. A key
determinant is the funding available for research \& development (R\&D). To
quantify the expenses in R\&D of a country we consider the fraction of
gross domestic product (GDP)
that is spent on R\&D. To get rid of economic inequalities in different
countries we consider the R\&D spending in terms of the purchasing power
parity (PPP). In Fig.~\ref{fig:gdp}A, we plot the number of citations
$N_{\mathrm{Cite}}$ against the R\&D expenditure and find that it scales
linearly with funding. Such correlation is not surprising, but the
scaling exponent is non-trivial. It suggests that it is not possible to
perform or contribute substantially unless there is a corresponding amount
of funding available for research. Moreover, the research contribution in
terms of citations also scales linearly with the number of researchers in that
country (Fig.~\ref{fig:gdp}B). This result is consistent with the fact that the R\&D expenditure is
correlated with the number of researchers.
The number of publications of a country also shows similar scaling against 
R\&D expenditure and number of researchers (Appendix Fig.~\ref{fig:gdpSI}). 

Finally as a measure of impact of a country's scientific output we consider
the average number of citations to the publications of that country. In
Fig.~\ref{fig:gdp}C we plot this number against the
average spending per researcher per year (R\&D expenditure divided by the
number of researchers).  The latter is not the average salary of
researchers in that country, as it includes other expenditures such as
infrastructure, bureaucracy, instruments, etc. This plot is much more
scattered than the previous plots and does not show any definite
correlation pattern. In order to identify groups of countries
that behave similarly or show similar characteristics we use the $k$-mean
clustering technique~\cite{Sculley10}. By using this clustering method with
$k=2$, we found that the countries can be classified into two groups, one
with average spending less than about 100,000 \$ per researcher per year and other
with average spending more than about 100,000 \$ (Fig.~S5).  Another
clustering methods also give qualitatively similar results. This separation
in two groups, distinguished by the average spending per researcher per
year (vertical line in the plot) also reveals another striking feature.
%Nevertheless, the plot reveals a striking feature. 
If the average spending is less than about 100,000 \$ (vertical line in the plot) per
researcher per year we see an increase in the average number of citations
with the spending. However if the average spending exceeds this limit, it
becomes scattered and independent of funding. This figure shows that very
rich countries like Kuwait and Luxembourg have high funding per researcher,
still the average number of citations per paper is low. Countries like
India, Brazil have high funding per researcher as well, but low average
number of cites; this might mean they are investing more on infrastructure.
Switzerland, Costa Rica, Panama, Germany, Austria, Netherlands, United
States have high spending per researcher and their average number of
citations is also high. If we
display the number of cites per paper averaged over all countries
(horizontal line), we see that there are no countries in the top left
quadrant, i.e. it is not possible to do better than the world's average
unless there is sufficient spending. Additional measures of a country's
research performance and corresponding rankings are reported in the
Appendix Table~\ref{tab:contribution}. 

\section{Discussion}
Our thorough analysis of the world citation and collaboration networks has
revealed that the effects of geography on the dynamics of science are relevant,
despite the recent advances in communication and transportation. The
occurrence of gravity laws for both citation and collaboration implies a
preference by scientists to interact with peers in their geographic areas.
However, long-distance interactions are not rare, as the interaction
strength and probability are characterized by power law decays. Our work
follows similar findings in mobile phone
communication~\cite{Lambiotte08,Krings09}, social
media~\cite{LibenNowell05} and international trade~\cite{Kaluza10},
reinforcing the belief that gravity laws hold in several different
contexts, and that scientific interactions are not exceptional from this
point of view. Thus, the gravity law is a fundamental
relationship holding also in human dynamics.

Citation and collaboration streams between distinct locations are strongly
correlated, with an approximately linear relation. An
increase in the number of collaborations between two cities is then expected to
be followed by a proportional increase in the flow of citations between the
cities. This is justified from the fact the people/groups working in
similar fields and subject area are more likely to cite as well as
collaborate with each other, and also suggests a natural bias towards
self-citation, of which we have provided strong quantitative evidence.

From the point of view of
scientific impact, it pays off for a team to put together several
institutions with a strong international participation. 
While part of this effect could be justified by the fact that having people
from different locations 
facilitates the circulation of a work, which then becomes more visible and
susceptible to be cited, the trend indicates that it is more likely
to produce high quality work through international collaborations. It would
be valuable to be able to disentangle the impact due to social networking
from that due to the quality of the paper. Our findings pave the way for
the first quantitative assessment of this issue.
As a consequence,
we expect to observe an increasing tendency to form large teams with
members of many different countries in the future. 

We also disclose a striking effect in the relationship between the national
expenditure per researcher and the impact of the scientific output of a
country. If the average spending per researcher per year is low,
it is impossible for a country to do better than the world average, in
terms of the average number of cites per paper. So there is a minimal
funding quota that needs to be exceeded if a country wishes to have a
scientific output of high average quality. Exceeding the threshold,
however, does not guarantee success.
This suggest that in science money acts as a kind of threshold
%We have also found that in science money acts as a kind of threshold
motivator: if one does not pay people enough they will not be motivated and
the outcomes of the research are poor; if people are paid sufficiently to
take the issue of money off the table, internationally competitive findings
are within reach. On the other hand, for conceptual and creative tasks,
paying more than a certain threshold does not necessarily increase the
output~\cite{Adams66,Alderfer72,Deci85}.
Further, our analysis reveals that at the country level funding has a
positive linear impact on the research output both in terms of number of
publications as well as citations.
Thus, it is not possible for a country to increase its research output
  substantially without a sizeable increase in investments.  

In the future we plan to study the role 
of cities' population, in particular on the
distributions of citation and collaboration strengths along with their
flows. It is well known that most characteristics of cities are
strongly correlated to the size of their populations~\cite{Bettencourt07}.
Furthermore, an analysis of the evolution of the world citation and
collaboration networks would show how the spatial
dimension of science dynamics has been affected by the progress of
technology, internationalization and extreme events (e.g. wars,
economic crises). 
This way one could infer how the scientific landscape has been shaping up in
the last decades and how is it possible to 
create more efficient partnerships, via dedicated funding
programs at the national and/or international level, and consequently
a more productive and successful scholarly world.

\section{Methods}
\subsection{Data description}
We have analyzed all publications (articles,
reviews and editorial comments) written in English from 2003 till the
end of 2010 included in the database of the Institute for
Scientific Information (ISI) Web of Science. For each publication we
extract the affiliations of the authors and the corresponding citations to
that publication. We parsed the affiliations of all publications and have
determined the geographic location at the city and country level. If
there are multiple affiliations listed in a publication, the latter is
associated with all represented cities and countries.  After obtaining the
locations we use the publicly available resources ({\tt
www.wikipedia.org} and {\tt maps.google.com}) to determine their
coordinates (latitude and longitude). 
Our dataset consists of 8,094,948 publications which have received 62,105,592
citations during the period 2003-2010. We were able to extract the
geographical information from 8,092,314 publications. Affiliations refer to
226 countries and 37,750 cities. In order to get rid of anomalies
due to any misclassification, we have only consider those places that have
appeared in at least 5 publications during the period 2003-2010.  This
cutoff led us to 18,199 cities, producing 99.8\% of the total publications
and receiving 99.9\% of total citations.

Country level information regarding expenditures for research and
development (R\&D) in terms of purchasing power parity (PPP) and number of
researchers in R\&D are obtained from the World Bank Data ({\tt
databank.worldbank.org}) for each year between 2003 till 2010. By
aggregating these yearly datasets we determine the average of each of the
above quantities for the period 2003-2010. The data of expenditure for R\&D
is available for 102 countries, the numbers of researchers 
for 89 countries and for 77 countries both informations are available.
Further details can be found in the Appendix.

\subsection{Network construction}
We have analyzed the data at the country and the
city level. As the publications and their affiliations form a bipartite
graph, we construct the collaboration network between countries (cities) by
projecting it onto the space of affiliations. In this collaboration network
individual countries (cities) act as nodes, and links between them
indicate that they have appeared in the same publication. If a
paper is written by authors with $n$ affiliations, 
we put $\frac{1}{2}n\times (n-1)$ undirected links
between each possible pair of collaborating countries (cities), with every
link having weight $\frac{2}{n \times (n-1)}$. The total weight between any pair
of nodes is the sum of all the weights over all the publications in the
dataset. If there is a single affiliation in a publication then we put
a self-link with weight 1. 

In the citation network between countries (cities) nodes are papers which
are linked if one paper cites the other. If a paper
written by authors with $n$ affiliations cites a paper written by
authors with $m$ affiliations we put
$n\times m$ directed connections from each of the $n$ citing countries
(cities) to each of the $m$ cited countries (cities), every link
having weight $1/(nm)$. The total weight of a directed link between two
countries (cities) is the sum of all the weights over all the citations in
the dataset. Since there can be multiple affiliations from the same country (city) in
a publication, there are self-loops both in the world citation and in
the world collaboration networks.

\subsection{Great-circle distance}
The geodesic or the great-circle distance is the shortest distance between
any two points on the earth measured along a path on the surface of the
earth. Given the latitudes and longitudes of two points, we have used the
Haversine formula to calculate the great-circle distance between
them~\cite{Sinnott84}. In these calculations, we considered the earth's
radius to be 6372.8 KM.

\begin{acknowledgments}
Financial support from EU’s FP7 FET-Open to ICTeCollective Project
No. 238597, and from the Academy of Finland, the
Finnish Center of Excellence program 2006-2011, Project No.
129670 are gratefully acknowledged.
Certain data included herein are derived from the Science Citation Index
Expanded, Social Science Citation Index and Arts \& Humanities Citation
Index, prepared by Thomson Reuters, Philadelphia, Pennsylvania, USA, 
Copyright Thomson Reuters, 2011
\end{acknowledgments}

\bibliography{mappingOfScience}

\appendix
\numberwithin{table}{section}
\numberwithin{figure}{section}

\section{Materials and methods}
\label{sec:AppendixMethod}
\subsection{Data description}
For our study we used all publications in English in the databases of
Science Citation Index Expanded, Social Sciences Citation Index, and Arts
\& Humanities Citation Index for the years 2003-2010. The database of the
Institute for Scientific Information (ISI) Web of Science also includes
publications in other major languages, but consists of a relatively small
number of items, accounting for $<5\%$ of total publications. For each
publication in the database, we have the name of the journal in which it is
published, the volume and page number of the publication, its year of
publication, the names of the authors, the list of their affiliations and its
references and other additional information. We used the list of references
to construct the network of citations between papers. For each publication
we extracted the city and country of authors’ institutions from the
affiliation data. Whenever a publication has several authors, it is counted
and assigned to each location.  Note that we only have the list of authors
and the list of affiliations for each paper, however there is no
corresponding match between these two lists and hence the individual level
author affiliation can not be used in our study.
Further although the affiliations are being recorded with increasing
consistency, their use still poses major challenges in uniquely and
accurately identifying them. For this reason, we parsed the affiliations of
all publications and have determined the geographic location only at the
city and country level. We also we use the publicly available resources
({\tt www.wikipedia.org} and {\tt maps.google.com}) to disambiguate the
names of the places in case there are multiple name variation, typos and
name changes during the time period of study.

\subsection{GDP}
The gross domestic product (GDP) is the value of all final goods and
services produced within a nation in a given year and is the primary
indicators used to gauge the health and size of a country's economy.  We
consider the average GDP (in US dollars) of a country during 2003-2010.  A nation's
GDP at purchasing power parity (PPP) exchange rates is the sum value of all
goods and services produced in the country valued at prices prevailing in
the United States. This is the measure most economists prefer when looking
at per-capita welfare and when comparing living conditions or use of
resources across countries.  

\subsection{R\&D spending} 
Expenditures for research and development are current and capital
expenditures (both public and private) on creative work undertaken
systematically to increase knowledge, including knowledge of humanity,
culture, and society, and the use of knowledge for new applications. R\&D
covers basic research, applied research, and experimental development.

\subsection{Number of researcher}
Researchers in R\&D are professionals engaged in the conception or creation
of new knowledge, products, processes, methods, or systems and in the
management of the projects concerned. Postgraduate PhD students engaged in
R\&D are included.

\subsection{Statistics}

To fit the data and calculate different estimates we use the following
methods:

{\bf Estimation of standard errors.} Bootstrapping is a distribution-free re-sampling method used to
estimate the parameters of interest from the empirical data. We have used
this method in order to calculate the standard error of the mean. Let
$x_1,x_2,\dots, x_n$ be the dataset with mean $\bar{x}$. The standard error
is then calculated as follows~\cite{Efron93}:
(i) Draw $N$ samples each of size $n$ with replacement from the original
data.
(ii) For each of the $N$ samples calculate the sample mean
$\hat{x_1},...,\hat{x_N}$
(iii) The standard error is then given by,
%\begin{equation}
  $SEE(\bar{x})=\sqrt{\frac{1}{N-1}\sum_{i=1}^{N} (\hat{x_i} -
  \bar{\hat{x_i}})^2},
$
%\end{equation}
where  $\bar{\hat{x_i}}=N^{-1}\sum_i^N \hat{x_i}$ is the mean of the $N$
bootstrap sample. In this study we have used $10^4$ bootstrapped samples, i.e.,
$N=10^4$.

{\bf Estimation of significance difference.}
The above bootstrapping procedure however does not tell whether the difference in the means of two distributions is
significant or not. In this case the re-sampling has to be performed
according to an appropriate null hypothesis, whereas for standard errors the
re-sampling procedure was unrestricted.

Let us consider two independent samples $x_1, \dots, x_n$ and $y_1,\dots,
y_m$, and suppose that we are interested in the difference in the
population means, $\delta = \bar{x}-\bar{y}$. Consider that the null hypothesis is
$H_0: \bar{x}-\bar{y} = 0$. We create the bootstrap sample by choosing $n$
elements without replacement from the pooled set $x_1, \dots, x_n,
y_1,\dots, y_m$. The remaining $m$ elements constitute the other sample.
We then calculate the mean of both these samples and determine the
difference between them, say $\hat{\delta}_i=\bar{\hat{x}}_i-\bar{\hat{y}}_i$. In
analogous fashion $N$ re-samples are made, and the bootstrap $p$ value is
defined as $p=\frac{(\#(\hat{\delta}_i \geq \delta,~\forall i)) +1}{N+1}$.
In this study we have used $10^4-1$ bootstrapped samples.

{\bf Power-law exponent.}
We use maximum likelihood techniques to estimate the scaling exponent of power
law distributions~\cite{Clauset09}.

{\bf Regression Coefficient.}
We used the linear regression analysis to study the relationship between
the corresponding variables. We determine the regression coefficient using
the ordinary least squares. The error term of the regression coefficient
represents the standard error of the estimate.

\subsection{Map construction}
Statistical data with embedded geographical information can be visualized
with standard maps which are color coded by region. However these maps are
sometimes hard to interpret as the statistical measures are often
correlated with the other indicators. We have used a diffusion-based method
to create different density-equalizing maps~\cite{Gastner04}. In this
method we start with an inhomogeneous distribution of the research
contribution (in terms of citations, say) and let the diffusion process
evolve until a homogeneous equilibrium state is reached: the displacements
are then reinterpreted to generate the cartogram. 

\section{Results}
\begin{figure*}[tbh!]
  \centering
  {\includegraphics[width=0.90\linewidth]{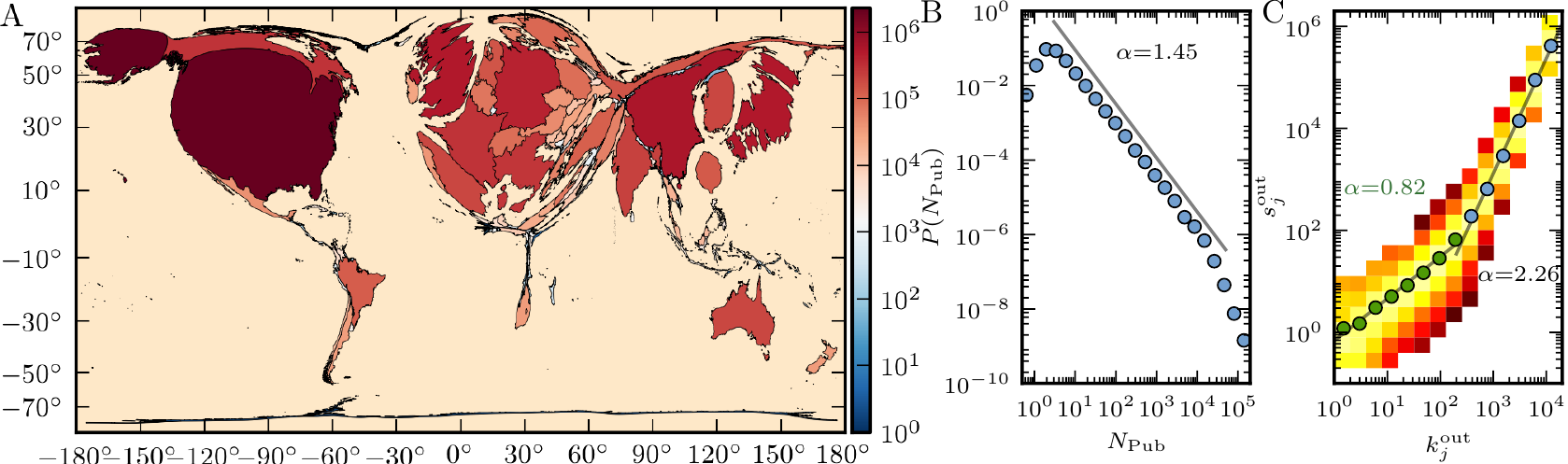}}
\caption{Research contribution in terms of number of publications. (A) Map
  of the country's research contribution, where the area of each country is
  scaled and deformed according to its number of publications. (B) The
  probability distribution function of the research contribution of cities
  in terms of their number of publications. The dashed line shows a power
  law scaling behavior with exponent $1.45\pm0.01$. (C) Node
    out-strength against its out-degree for city citation network. There
    are two distinct power law scaling regions, with scaling exponents
  $0.82\pm0.04$ and $2.26\pm0.07$ for low and high degree ($>200$) nodes,
respectively.}
\label{fig:citationSI}
\end{figure*}
\begin{figure*}[tbh!]
  \centering
  {\includegraphics[width=0.90\linewidth]{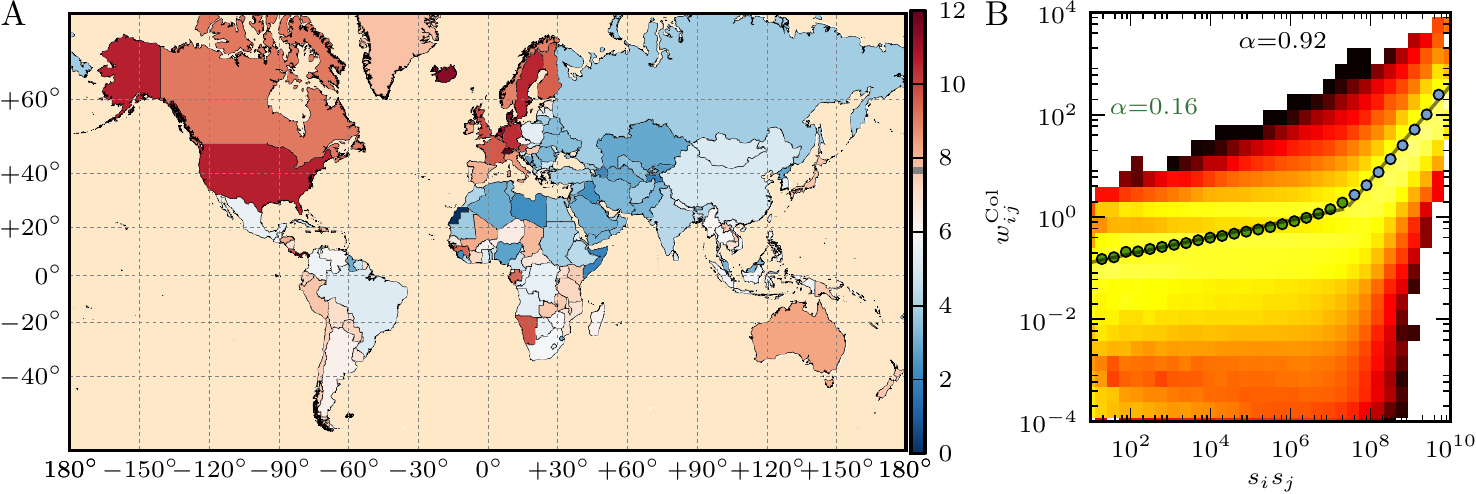}}
\caption{(A) Average number of citations of each country. World map where countries
  are color coded based on
the average number of citations per publication. Most countries stay
below the world's average of 7.67.(B) Weight of the links against the product of
the strengths of the endpoints in the collaboration network of cities. There are two different scaling regions
($2\times10^7$), with exponents $0.16\pm0.01$ and $0.92\pm0.03$.
}
\label{fig:averageCitationSI}
\end{figure*}
We consider the research contribution of each country in terms of the
number of publications $N_{\mathrm{Pub}}$, normalized by the number of
participating countries in that publication.  To visualize the results, we
create a cartogram in which the geographic regions are deformed and
rescaled in proportion to their relative research
contribution~\cite{Gastner04}. We observed that the contribution of
different countries in terms of publications is heterogeneous and varies
over 6 order of magnitude. Fig.~\ref{fig:citationSI}A shows that North
America (32.4\%), Europe(33.7\%) and Asia(27.4\%) have prominent
contribution in terms of the number of publications. On the other hand,
Africa, South America and Oceania contribute less than 7\% of world's
publications. Table.~\ref{tab:contribution} shows the contribution, number
of countries and cities in each continents. It also indicates the
statistics of the top countries of each continent. It is evident
that the United States are the leading country in the world both in terms of
publications and citations to them. It is followed by China, United
Kingdom, Japan, and Germany in terms of publications, whereas in terms of
citations it is followed by United Kingdom, Germany, Japan, and China. We
indicate the fraction of total publications $f_{\mathrm{Pub}}$, the fraction of total
citations received $f_{\mathrm{Cite}}$ and the average number of citations per paper, for
countries that received more than 0.005\% of world citations. Countries
are listed in decreasing order of the fraction of total citations
received. The superscripts in  $f_{\mathrm{Pub}}$ and
$f_{\mathrm{Cite}}$ indicate the world ranking of that country
according to the numbers of publications and citations, respectively. We then
consider the contribution in terms of the number of publications at the level of cities. In
Fig.~\ref{fig:citationSI}B we plot the probability distribution of the
cities' contributions in terms of their publications and observed that it 
follows a power law scaling behavior with exponent $1.45\pm0.01$. 
By plotting the out-degree against the out-strength, we find that there is power law scaling
behavior with $\langle s^{\mathrm{out}} \rangle (k^{\mathrm{out}}) \propto
(k^{\mathrm{out}})^{\alpha}$
(Fig.~\ref{fig:citationSI}C). However, there are two distinct scaling
regimes: for nodes with small $k_i^{\mathrm{in}}$ ($<200$) the exponent is
$\alpha=0.82\pm04$, while for large $k_i^{\mathrm{out}}$ ($\geq200$) the
exponent is $\alpha=2.26\pm0.07$. The super-linear behavior suggests that
stronger links are more frequently connected to high out-degree nodes.

\begin{figure}[tb]
  \centering
  \includegraphics[width=0.90\linewidth]{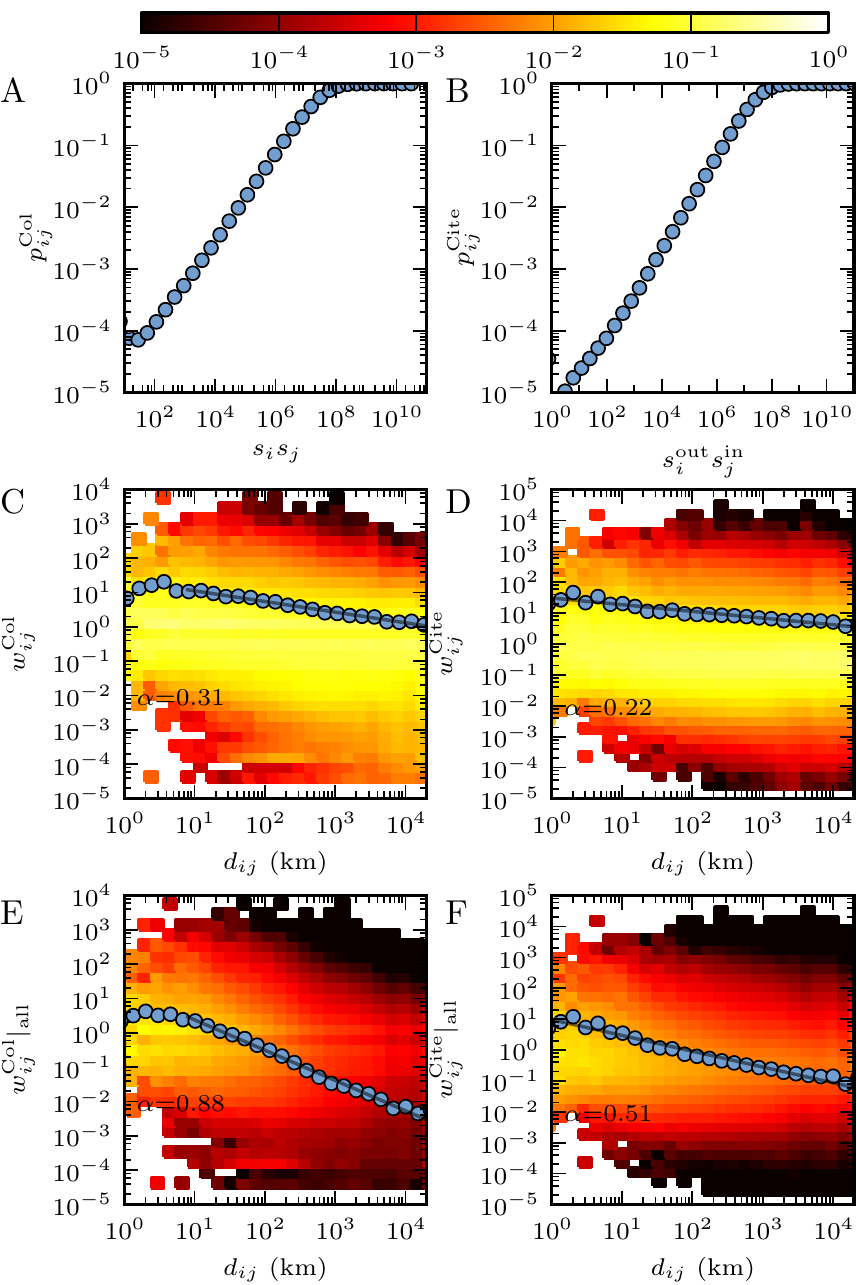}
\caption{Gravity law in the world collaboration and citation networks.(A) Variation
  of the probability of existence of a link between two nodes as a
  function of the product of their strengths in the (A) collaboration network
  and (B) citation network of cities. Variation of the average
  link weight against the distance between the cities in the (C)
  collaboration network and (D) citation network.  For each distance the
  average ratio is also shown. In this case only the existing links are
  considered while calculating the averages. The solid line indicates a
  power law behavior with exponent $\alpha=0.31\pm0.01$ and
  $0.22\pm0.01$ respectively.  Variation of the average link weight against the distance between the
  cities in the (E) collaboration network and (F) citation network.  For each distance the average ratio
  is also shown. In this case all possible node pairs are considered in
  order to calculate the average, i.e.,  links that do not exist are
  considered with weight 0. The solid line indicates a power law behavior
  with exponent $\alpha=0.88\pm0.02$ and $0.51\pm0.02$, respectively.
}
\label{fig:gravityLawSI}
\end{figure}
\begin{figure}[tb]
  \centering
  {\includegraphics[width=0.95\linewidth]{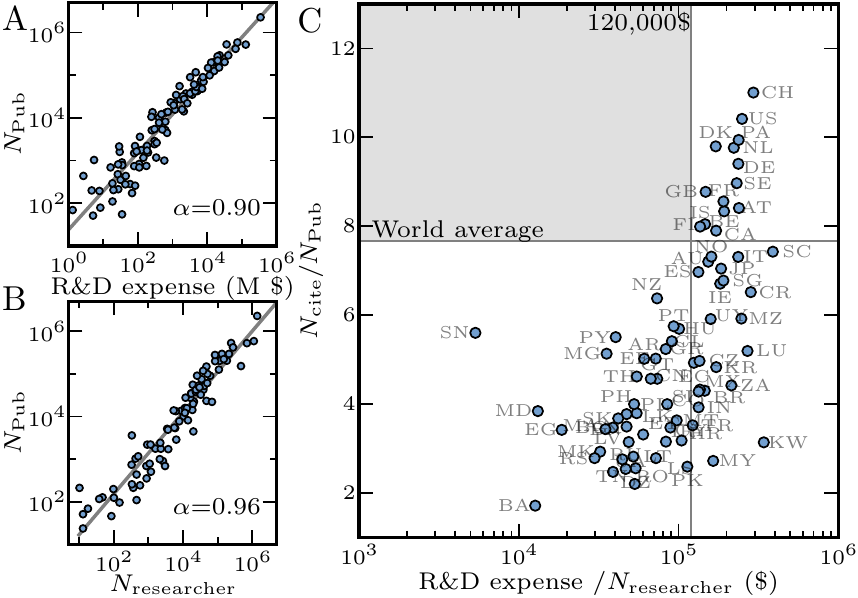}}
\caption{Relation between research contribution in terms of number of publications and
  funding. Country's number of publications against the (A) expenditure in
  research and development (in million dollars, and purchasing power
  parity), (B) number of researchers in that country. The solid line
  indicates a scaling with exponent $0.90\pm0.03$ and $0.96\pm0.03$,
  respectively. (C) The plot of average spending per researcher against the
  average number of citation per paper of that country. The average
  number of 
  citations is now defined as
  the ratio of the normalized number of citations and normalized number
  of publications (see text). The horizontal line indicates world
  average, 
the vertical line indicates the spending of 120\,000 \$
  per researcher.}
\label{fig:gdpSI}
\end{figure}

Next we consider the average number of citations per paper of each country and plot it on a
colorpleth map~(Fig.~\ref{fig:averageCitationSI}A). For calculating the
average citation of a country we consider all its publications and count
the total number of citations to all these articles during the period of
2003-2010.  In the case where a publication has multiple affiliations from
different countries, it is counted multiple times for the countries' averages,
once for each of the affiliated countries.  In
Table~\ref{tab:contribution}, we have also given the average number of
citations per paper of
the top countries in each continent. The world average is 7.67. United States, Canada, Australia and
most of the European countries have average number of citations larger than the world
average.  In Europe Switzerland leads the table, followed by
Denmark and Netherlands. In contrast most of the countries from Asia
stay below the world average, the only exception being Israel.  Most of the
countries in Africa and South America are below the
world average as well. Other notable countries 
are Bermuda (16.97$\pm$5.95), Gambia (16.17$\pm$3.10), Panama
(12.41$\pm$ 0.68), Iceland (11.43$\pm$ 0.71), Seychelles (11.11$\pm$ 2.40),
Guinea-Bissau (10.10$\pm$ 0.97), Costa Rica (9.82$\pm$ 0.93), and Austria
(9.75$\pm$ 0.09). 
For the collaboration network of cities we plot the weight of the
links against the product of the strengths of the connecting nodes,
expressing the expected weight of random collaborations
(Fig.~\ref{fig:averageCitationSI}B). As for citations we find that
$w_{ij}^{\mathrm{Col}} \propto (s_is_j)^{\alpha}$, with two different
scaling exponents. If $s_is_j<2\times10^7$, $\alpha=0.16\pm0.01$
($R=0.11$), whereas if $s_is_j>2\times10^7$ $\alpha=0.92\pm0.03$
($R=1.18$). 

In Fig.~\ref{fig:gravityLawSI}A,B we plot the probability of existence of
a link as a function of the product of strength of the end-points of
the link. We
found that as the product increases, both in the collaboration and the
citation network the probability of link existence increases, as expected. In
Fig.~\ref{fig:gravityLawSI}C,D we show the variation of the link weight
against the distance between the end-points. We found that both in the
collaboration and the citation network on the average the link weight
decreases as a power-law with exponent $0.31\pm0.01$ and $0.22\pm0.01$,
respectively. In this figure, while calculating the averages we have only
considered the existing links between nodes. However, in the main text we
have seen that the probability of link existence also decreases with
distance. If we take this information while calculating the averages, i.e.,
we consider the non-existent links by assigning weight zero to them, we found that in both
the collaboration and the citation network, the average link weight
decreases with distance as a power law, with exponent $0.88\pm0.01$ and
$0.51\pm0.01$, respectively (Fig.~\ref{fig:gravityLawSI}E,F). Note that
this property is different from what has been observed in the mobile phone
communication network, where it was shown that the weight of the existing
links are independent of the distance, whereas the overall link weight
decrease as a result of decreasing probability of having a link as the
distance increases~\cite{Krings09}. 

In the main paper we have considered the research performance of each
country based on the  number of citations. In addition, here we consider
the performance of a country based on its number of publications. As before,
in Fig.~\ref{fig:gdpSI}A, we plot the research contribution in terms of the
number of publications $N_{\mathrm{Pub}}$ against the countries' R\&D
expenditure in terms of purchasing power parity (PPP). We found that this
indicator also scale almost linearly with the spending. We next consider
the dependence of research performance on the number of researchers in that
country (Fig.~\ref{fig:gdpSI}B). The research contribution in terms of publications also scale
linearly with the number of researchers in that county.

Finally as a measure of the average publication quality of a country we
consider the ratio of the normalized number of citations and
normalized number of publications of
that country. This is an alternative measure of the average number of
citations per paper we mentioned above, which is
not normalized by the number of authors in a paper. In the previous measure
each publication from a country (independently of the number of
participating countries) gets equal weight while calculating
the average. In this other measure, if there are $n$ countries in a
publication, each country would
get $1/n$ as credit for that publication, so that publication would
give a lower contribution to the average number of cites per paper
than before.
In Fig.~\ref{fig:gdpSI}C we plot the new quantity against the average
spending per researcher of the country
(R\&D expenditure divided by the number of researchers). Although this plot
is similar to the one in the main paper, there are certain
differences in the average number of citations of some countries. For example, Italy,
Spain, Norway are now below the world average. This
means that the publications from these countries with international
collaborators contribute significantly to the average impact of their
scientific production.

\begin{figure}[tb]
  \centering
  {\includegraphics[width=0.49\linewidth]{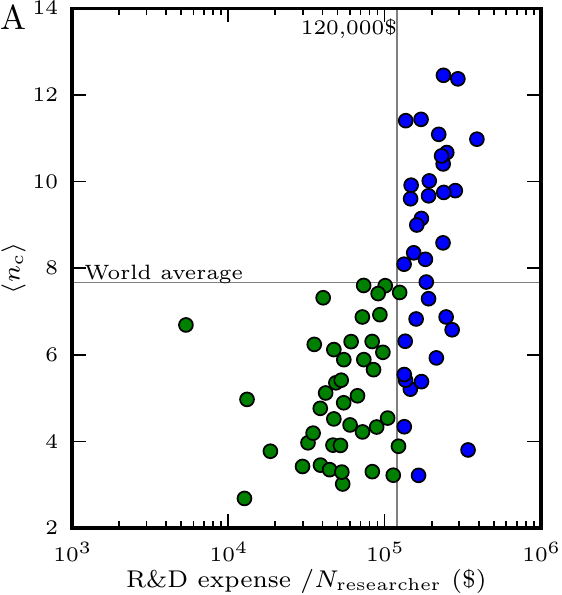}}
  {\includegraphics[width=0.49\linewidth]{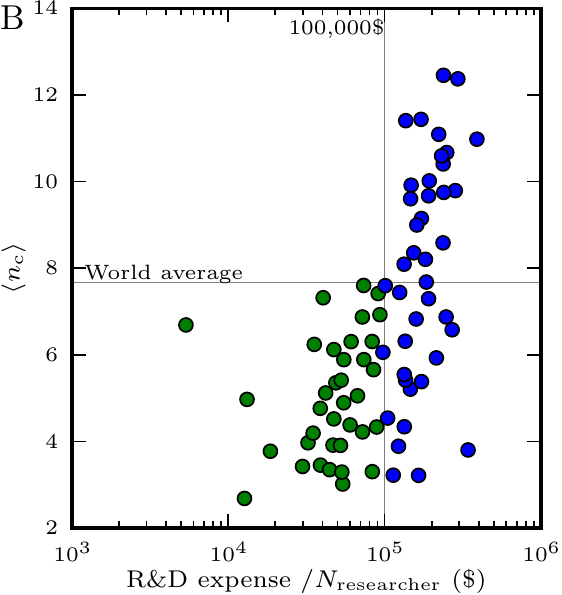}}
\caption{Data clustering. (A) Decomposition obtained using $k$-mean
  clustering with $k=2$. (B) Decomposition obtained using mean shift
  clustering. Each cluster is indicated by a color.
}
\label{fig:gdpClusteringSI}
\end{figure}
In order to check whether the countries in Fig. 5C can be categorized
  into different groups based on the average spending per researcher and
  the average number of citations, we used two different clustering methods.
  The $k$-means clustering technique~\cite{Sculley10} partitions the data
  into $k$-mutually exclusive clusters. The aim here is to determine
  whether there are inherent clusters in Fig.5C and Fig~\ref{fig:gdpSI}C. For
  the $k$-means clustering method we need to specify the number $k$ of
  clusters 
  before starting the clustering process.  The method consists in the
  minimization of an objective function expressing the sum of
  square distances between each data point and its {\it centroid},
  i.e. a geometrical point whose position is also consistently
  determined by the minimization procedure: each centroid corresponds
  to one cluster.
  We can follow a procedure to minimize the objective function iteratively
  by finding a new set of cluster centroids that can lower the value of the
  objective function at each iteration. On using this method with $k=2$, we
  found that the countries can be classified into two groups, one with
  average spending less than about 120,000 \$ per researcher per year and other
  with average spending more than about 120,000 \$
  (Fig.~\ref{fig:gdpClusteringSI}).
  We also use a different method, the mean shift clustering
algorithm~\cite{Comaniciu02} to determine the clusters in the data in
Fig.5. This is a nonparametric clustering technique and does not require
prior knowledge of the number of clusters. The mean-shift algorithm seeks
local maxima of density of points in the feature space. This method also
detects two different clusters, one with average spending less than about
100,000 \$ per researcher per year and the other with average spending more
than about 100,000 \$. Thus, these two methods give slightly different
thresholds however the results are qualitatively similar.

%\clearpage
\LTcapwidth=\textwidth
\begin{longtable*}{lccccccccc}
\caption{Research contribution of different continents and their top
  countries. The number of countries and cities in each continent are
  indicated by $N_{\mathrm{Countries}}$ and $N_{\mathrm{Cities}}$,
  respectively. Fraction of publications $f_{\mathrm{Pub}}$, fraction of
  citations received $f_{\mathrm{Cite}}$ and the average number of
  citations per paper of each
  continent is also indicated.  For top countries in each continent we
  list the fraction of publications $f_{\mathrm{Pub}}$, fraction of
  citations received $f_{\mathrm{Cite}}$, the average number of
  citations per paper. The
  superscript indicates the countries' rank in the world in terms of
  number of publications and number of citations. Only countries that
  receive more than 0.005\% of all citations are shown. 
}\\
\noalign{\smallskip} \hline \noalign{\smallskip}
Continent & $N_{\mathrm{Countries}}$ & $N_{\mathrm{Cities}}$ &
$f_{\mathrm{Pub}}$ & $f_{\mathrm{Cite}}$ & Avg. Cites & 
           Country & $f_{\mathrm{Pub}}$ & $f_{\mathrm{Cite}}$ & Avg. Cites \\ 
        & & & (in \%) & (in \%) & & name & (in \%) & (in \%) & \\ 
  \endfirsthead
  \multicolumn{10}{c}%
    {\tablename\ \thetable\ -- \textit{Continued from previous page}} \\
\noalign{\smallskip} \hline \noalign{\smallskip}
Continent & $N_{\mathrm{Countries}}$ & $N_{\mathrm{Cities}}$ &
$f_{\mathrm{Pub}}$ & $f_{\mathrm{Cite}}$ & Avg. Cites & 
           Country & $f_{\mathrm{Pub}}$ & $f_{\mathrm{Cite}}$ & Avg. Cites \\ 
        & & & (in \%) & (in \%) & & name & (in \%) & (in \%) & \\ 
\noalign{\smallskip} \hline \noalign{\smallskip}
    \endhead

    %\noalign{\smallskip} \hline \noalign{\smallskip}
    \multicolumn{10}{r}{\textit{Continued on next page}} \\
    \endfoot
    \midrule
    \endlastfoot
\noalign{\smallskip} \hline \noalign{\smallskip}
\multirow{18}*{Africa}        & \multirow{18}*{57} &  \multirow{18}*{749}  & \multirow{18}*{1.32}  & \multirow{18}*{0.65}  & \multirow{18}*{5.00$\pm$0.05} & South Africa & $0.430^{33}$ & $0.248^{37}$ & 5.92$\pm$0.08 \\
                             &    &       &       &       &               &        Egypt &   $0.286^{38}$ & $0.128^{40}$ & 3.78$\pm$0.05\\     
                             &    &       &       &       &               &      Tunisia &  $0.100^{52}$ & $0.036^{52}$ & 3.33$\pm$0.13\\     
                             &    &       &       &       &               &      Nigeria &  $0.126^{50}$ & $0.031^{56}$ & 2.82$\pm$0.25\\     
                             &    &       &       &       &               &        Kenya &  $0.038^{65}$ & $0.028^{58}$ & 7.55$\pm$0.29\\     
                             &    &       &       &       &               &      Morocco &  $0.055^{60}$ & $0.025^{60}$ & 4.20$\pm$0.10\\     
                             &    &       &       &       &               &      Algeria &  $0.067^{54}$ & $0.023^{62}$ & 3.01$\pm$0.08\\     
                             &    &       &       &       &               &     Tanzania &  $0.019^{83}$ & $0.014^{74}$ & 7.27$\pm$0.29\\     
                             &    &       &       &       &               &       Uganda &  $0.018^{85}$ & $0.014^{75}$ & 7.04$\pm$0.27\\     
                             &    &       &       &       &               &     Cameroon &  $0.021^{77}$ & $0.010^{80}$ & 4.72$\pm$0.18\\     
                             &    &       &       &       &               &     Ethiopia &  $0.022^{75}$ & $0.009^{81}$ & 4.36$\pm$0.17\\     
                             &    &       &       &       &               &        Ghana &  $0.016^{86}$ & $0.008^{85}$ & 5.29$\pm$0.21\\     
                             &    &       &       &       &               &     Zimbabwe &  $0.011^{95}$ & $0.007^{87}$ & 5.92$\pm$0.28\\     
                             &    &       &       &       &               &       Malawi &  $0.009^{103}$ & $0.006^{88}$ & 7.11$\pm$0.32\\     
                             &    &       &       &       &               &     Senegal  &  $0.008^{104}$ & $0.006^{90}$ & 6.72$\pm$0.29\\     
                             &    &       &       &       &               &     Botswana &  $0.010^{97}$ & $0.005^{95}$ & 5.46$\pm$0.54\\     
                             &    &       &       &       &               &       Gambia &  $0.003^{128}$ & $0.005^{96}$ & 15.87$\pm$3.02\\     
                             &    &       &       &       &               & Cote d'Ivoire & $0.006^{107}$ & $0.005^{97}$ & 7.20$\pm$0.41\\ \noalign{\smallskip} \hline \noalign{\smallskip} 
\multirow{14}*{Asia}         & \multirow{14}*{49} &  \multirow{14}*{3853} & \multirow{14}*{27.36} & \multirow{14}*{17.71} & \multirow{14}*{5.58$\pm$0.01} & Japan   &   $6.457^{4}$ & $5.939^{4}$ & 7.68$\pm$0.03    \\ 
                             &    &       &       &       &               & China        &  $7.216^{2}$ & $4.304^{5}$ & 5.05$\pm$0.02    \\ 
                             &    &       &       &       &               & South Korea  &  $2.509^{10}$ & $1.582^{13}$ & 5.38$\pm$0.04  \\ 
                             &    &       &       &       &               & India        &  $2.727^{9}$ & $1.398^{15}$ & 4.35$\pm$0.03   \\ 
                             &    &       &       &       &               & Taiwan       &  $1.671^{15}$ & $1.037^{16}$ & 5.19$\pm$0.04  \\ 
                             &    &       &       &       &               & Israel       &  $0.863^{22}$ & $0.837^{20}$ & 8.86$\pm$0.10  \\ 
                             &    &       &       &       &               & Turkey       &  $1.450^{17}$ & $0.667^{22}$ & 3.89$\pm$0.03  \\ 
                             &    &       &       &       &               & Russia       &  $1.875^{13}$ & $0.622^{24}$ & 3.92$\pm$0.05  \\ 
                             &    &       &       &       &               &  Singapore   &  $0.521^{29}$ & $0.461^{28}$ & 7.29$\pm$0.08  \\ 
                             &    &       &       &       &               &       Iran   &  $0.747^{23}$ & $0.308^{31}$ & 3.31$\pm$0.03  \\ 
                             &    &       &       &       &               &   Thailand   &  $0.244^{41}$ & $0.147^{38}$ & 5.88$\pm$0.17  \\ 
                             &    &       &       &       &               &   Malaysia   &  $0.195^{42}$ & $0.069^{45}$ & 3.22$\pm$0.07  \\ 
                             &    &       &       &       &               &   Pakistan   &  $0.175^{44}$ & $0.059^{48}$ & 3.23$\pm$0.07  \\ 
                             &    &       &       &       &               & Saudi Arabia &  $0.131^{49}$ & $0.046^{50}$ & 3.12$\pm$0.07  \\ 
 \multirow{15}*{Asia}         & \multirow{15}*{49} &  \multirow{15}*{3853} & \multirow{15}*{27.36} & \multirow{15}*{17.71} & \multirow{15}*{5.58$\pm$0.01}          &     Jordan   &  $0.061^{58}$ & $0.023^{61}$ & 3.27$\pm$0.11  \\ 
                             &    &       &       &       &               &    Vietnam   &  $0.037^{68}$ & $0.020^{66}$ & 5.59$\pm$0.30  \\ 
                             &    &       &       &       &               &  Indonesia   &  $0.032^{70}$ & $0.019^{67}$ & 5.73$\pm$0.23  \\ 
                             &    &       &       &       &               &     Kuwait   &  $0.044^{61}$ & $0.018^{68}$ & 3.80$\pm$0.16  \\ 
                             &    &       &       &       &               & Bangladesh   &  $0.041^{63}$ & $0.018^{69}$ & 4.74$\pm$0.17  \\ 
                             &    &       &       &       &               &    Lebanon   &  $0.038^{66}$ & $0.018^{70}$ & 4.47$\pm$0.13  \\ 
                             &    &       &       &       &               & UAE          &  $0.041^{62}$ & $0.018^{71}$ & 4.03$\pm$0.14  \\ 
                             &    &       &       &       &               & Philippines  &  $0.035^{69}$ & $0.017^{72}$ & 6.20$\pm$0.30  \\ 
                             &    &       &       &       &               &     Cyprus   &  $0.027^{73}$ & $0.012^{76}$ & 4.31$\pm$0.15  \\ 
                             &    &       &       &       &               &  Sri Lanka   &  $0.022^{76}$ & $0.011^{79}$ & 4.90$\pm$0.18  \\ 
                             &    &       &       &       &               &    Armenia   &  $0.026^{74}$ & $0.009^{82}$ & 6.18$\pm$0.31  \\ 
                             &    &       &       &       &               &       Oman   &  $0.021^{79}$ & $0.008^{86}$ & 3.50$\pm$0.14  \\ 
                             &    &       &       &       &               &    Georgia   &  $0.020^{82}$ & $0.006^{89}$ & 3.94$\pm$0.20  \\ 
                             &    &       &       &       &               &      Nepal   &  $0.012^{92}$ & $0.006^{91}$ & 5.36$\pm$0.29  \\ 
                             &    &       &       &       &               & Uzbekistan   &  $0.020^{81}$ & $0.006^{92}$ & 3.50$\pm$0.17  \\ \noalign{\smallskip} \hline \noalign{\smallskip} %\pagebreak
 \multirow{33}*{Europe}       & \multirow{33}*{47} &  \multirow{33}*{6625} & \multirow{33}*{33.69} & \multirow{33}*{35.25} & \multirow{33}*{9.29$\pm$0.01}& United Kingdom & $6.509^{3}$ & $7.453^{2}$ & 9.91$\pm$0.04  \\ 
                             &    &       &       &       &               &       Germany &  $5.131^{5}$ & $6.299^{3}$ & 10.41$\pm$0.04   \\
                             &    &       &       &       &               &        France &  $3.611^{7}$ & $4.034^{6}$ & 9.67$\pm$0.04    \\
                             &    &       &       &       &               &         Italy &  $3.415^{8}$ & $3.258^{8}$ & 8.59$\pm$0.04    \\
                             &    &       &       &       &               &  Netherlands  &  $1.829^{14}$ & $2.331^{9}$ & 11.08$\pm$0.07  \\
                             &    &       &       &       &               &        Spain  &  $2.482^{11}$ & $2.258^{11}$ & 8.09$\pm$0.05  \\
                             &    &       &       &       &               &  Switzerland  &  $1.114^{19}$ & $1.600^{12}$ & 12.38$\pm$0.09 \\
                             &    &       &       &       &               &       Sweden  &  $1.227^{18}$ & $1.436^{14}$ & 10.59$\pm$0.09 \\
                             &    &       &       &       &               &      Belgium  &  $0.923^{21}$ & $1.004^{17}$ & 10.02$\pm$0.08 \\
                             &    &       &       &       &               &      Denmark  &  $0.655^{25}$ & $0.838^{19}$ & 11.45$\pm$0.12 \\
                             &    &       &       &       &               &      Finland  &  $0.640^{26}$ & $0.672^{21}$ & 9.59$\pm$0.10  \\
                             &    &       &       &       &               &      Austria  &  $0.595^{27}$ & $0.653^{23}$ & 9.75$\pm$0.11  \\
                             &    &       &       &       &               &       Poland  &  $1.110^{20}$ & $0.579^{25}$ & 5.43$\pm$0.05  \\
                             &    &       &       &       &               &       Norway  &  $0.506^{30}$ & $0.483^{26}$ & 8.98$\pm$0.10  \\
                             &    &       &       &       &               &       Greece  &  $0.684^{24}$ & $0.468^{27}$ & 6.30$\pm$0.06  \\
                             &    &       &       &       &               &     Portugal  &  $0.460^{32}$ & $0.345^{30}$ & 6.93$\pm$0.08  \\
                             &    &       &       &       &               &Czech Republic &  $0.464^{31}$ & $0.301^{32}$ & 6.31$\pm$0.08  \\
                             &    &       &       &       &               &       Ireland &  $0.340^{36}$ & $0.298^{33}$ & 8.20$\pm$0.16  \\
                             &    &       &       &       &               &       Hungary &  $0.338^{37}$ & $0.251^{36}$ & 7.61$\pm$0.13  \\
                             &    &       &       &       &               &      Slovenia &  $0.171^{45}$ & $0.096^{41}$ & 5.41$\pm$0.09  \\
                             &    &       &       &       &               &       Ukraine &  $0.271^{40}$ & $0.088^{42}$ & 3.46$\pm$0.07  \\
                             &    &       &       &       &               &       Romania &  $0.274^{39}$ & $0.079^{43}$ & 3.30$\pm$0.09  \\
                             &    &       &       &       &               &      Slovakia &  $0.150^{48}$ & $0.072^{44}$ & 5.12$\pm$0.12  \\
                             &    &       &       &       &               &       Croatia &  $0.164^{47}$ & $0.068^{46}$ & 4.53$\pm$0.12  \\
                             &    &       &       &       &               &        Serbia &  $0.167^{46}$ & $0.061^{47}$ & 3.42$\pm$0.07  \\
                             &    &       &       &       &               &      Bulgaria &  $0.125^{51}$ & $0.056^{49}$ & 4.76$\pm$0.09  \\
                             &    &       &       &       &               &       Estonia &  $0.063^{57}$ & $0.041^{51}$ & 6.91$\pm$0.21  \\
                             &    &       &       &       &               &     Lithuania &  $0.095^{53}$ & $0.035^{53}$ & 3.91$\pm$0.13  \\
                             &    &       &       &       &               &       Iceland &  $0.030^{71}$ & $0.031^{57}$ & 11.46$\pm$0.63 \\
                             &    &       &       &       &               &       Belarus &  $0.064^{56}$ & $0.020^{65}$ & 3.37$\pm$0.10  \\
                             &    &       &       &       &               &        Latvia &  $0.021^{78}$ & $0.009^{83}$ & 5.34$\pm$0.43  \\
                             &    &       &       &       &               &    Luxembourg &  $0.012^{91}$ & $0.008^{84}$ & 6.58$\pm$0.34  \\
                             &    &       &       &       &               &       Moldova &  $0.011^{96}$ & $0.006^{93}$ & 4.94$\pm$0.31  \\ \noalign{\smallskip} \hline \noalign{\smallskip}
\multirow{7}*{North America} & \multirow{7}*{37} &  \multirow{7}*{5346} & \multirow{7}*{32.40} & \multirow{7}*{42.33} &\multirow{7}*{10.36$\pm$0.02} & United States &  $28.116^{1}$ & $38.216^{1}$ & 10.67$\pm$0.02 \\
                             &    &       &       &       &               &       Canada&  $3.616^{6}$  & $3.728^{7}$  & 9.15$\pm$0.05  \\
                             &    &       &       &       &               &      Mexico &  $0.523^{28}$ & $0.292^{34}$ & 5.57$\pm$0.10  \\
                             &    &       &       &       &               & Puerto Rico &  $0.037^{67}$ & $0.028^{59}$ & 7.66$\pm$0.26  \\
                             &    &       &       &       &               &        Cuba &  $0.040^{64}$ & $0.022^{63}$ & 4.81$\pm$0.14  \\
                             &    &       &       &       &               &  Costa Rica &  $0.014^{88}$ & $0.012^{77}$ & 9.93$\pm$0.87  \\
                             &    &       &       &       &               &     Panama  &  $0.009^{102}$ & $0.012^{78}$ & 12.43$\pm$0.81 \\ \noalign{\smallskip} \hline \noalign{\smallskip}
\multirow{2}*{Oceania}       & \multirow{2}*{21} &  \multirow{2}*{844}  & \multirow{2}*{2.89}  & \multirow{2}*{2.67}  & \multirow{2}*{8.22$\pm$0.05} & Australia & $2.448^{12}$ & $2.301^{10}$ & 8.36$\pm$0.05 \\
                             &    &       &       &       &               & New Zealand & $0.425^{34}$ & $0.354^{29}$ & 7.60$\pm$0.10 \\ \noalign{\smallskip} \hline \noalign{\smallskip}
\multirow{8}*{South America} & \multirow{8}*{14} &  \multirow{8}*{782}  & \multirow{8}*{2.34}  & \multirow{8}*{1.39}  & \multirow{8}*{5.75$\pm$0.04} & Brazil & $1.551^{16}$ & $0.871^{18}$ & 5.21$\pm$0.04 \\
                             &    &       &       &       &               & Argentina& $0.399^{35}$ & $0.261^{35}$ & 6.31$\pm$0.10 \\
                             &    &       &       &       &               &     Chile& $0.193^{43}$ & $0.136^{39}$ & 7.42$\pm$0.16 \\
                             &    &       &       &       &               &  Colombia& $0.066^{55}$ & $0.034^{54}$ & 5.65$\pm$0.19 \\
                             &    &       &       &       &               & Venezuela& $0.060^{59}$ & $0.034^{55}$ & 6.11$\pm$0.28 \\
                             &    &       &       &       &               &   Uruguay& $0.027^{72}$ & $0.021^{64}$ & 6.81$\pm$0.21 \\
                             &    &       &       &       &               &      Peru& $0.019^{84}$ & $0.014^{73}$ & 7.68$\pm$0.30 \\
                             &    &       &       &       &               &  Ecuador & $0.008^{105}$ & $0.005^{94}$ & 7.39$\pm$0.37 \\
\noalign{\smallskip} \hline \noalign{\smallskip}
\label{tab:contribution}
\end{longtable*}

\end{document}